%% file: main_ASE_camera_ready.tex
\documentclass[10pt,conference]{IEEEtran}
\IEEEoverridecommandlockouts
\RequirePackage{fix-cm}

\usepackage{rotating}
\usepackage{threeparttable, tablefootnote}
\usepackage{enumerate}

\usepackage{tikz}

\definecolor{gray(x11gray)}{rgb}{0.75, 0.75, 0.75}

\newcommand*\circled[1]{%
\tikz[baseline=(C.base)]\node[draw,circle,inner sep=0.5pt](C) {#1};\!}

\usepackage{balance}
\usepackage{booktabs}
\usepackage{graphicx}
\usepackage{nicefrac}
\usepackage[T1]{fontenc}
\usepackage{soul}
\usepackage{xspace} 
\usepackage{color}
\usepackage{url}
\usepackage{booktabs}
\usepackage{multirow}
\usepackage{listliketab}
\usepackage{amssymb}
\usepackage{algorithmic}
\usepackage{booktabs}
\usepackage{comment}
\usepackage{amsmath}
\usepackage{pgfplots}
\usepackage{bbding}
\usepackage{listings}
\usepackage{float}
\usepackage{tabularx}
\usepackage{cite}
\usepackage{tcolorbox}
\usepackage[normalem]{ulem}
\usepackage{censor}
\usepackage{tikz}
\usepackage{amssymb} 
\usepackage{amsfonts}
\usepackage{xparse}
\usepackage[colorinlistoftodos]{todonotes}
\usepackage{amssymb}
\usepackage{mathtools, nccmath}
\usepackage{venndiagram}
\usepackage{balance}
\usepackage{subfigure}
\usepackage{esvect}
\usepackage{xurl}
\usepackage{pgfplotstable}
\usepackage{cite}
\usepackage{amsmath,amssymb,amsfonts}
\usepackage[sort&compress, numbers]{natbib}
\usepackage{algorithmic}
\usepackage{graphicx}
\usepackage{textcomp}
\usepackage{xcolor}
\usepackage{hyperref}
\usepackage{circledsteps}

\newlength\WIDTHOFBAR
\date{May 2022}

\newcommand\RqOne{RQ1}
\newcommand\RqTwo{RQ2}
\newcommand\RqThree{RQ3}
\newcommand\RqOneText{How often are failing CI jobs rechecked?}
\newcommand\RqTwoText{How often do CI outcomes change after a recheck?}
\newcommand\RqThreeText{How much overhead is generated by rechecking builds?}

\DeclareRobustCommand{\IEEEauthorrefmark}[1]{\smash{\textsuperscript{\footnotesize #1}}}

\input{comments}
\begin{document}

\title{Repeated Builds During Code Review: \\An Empirical Study of the OpenStack Community}

\author{
\IEEEauthorblockN{
Rungroj~Maipradit\IEEEauthorrefmark{1}, Dong~Wang\IEEEauthorrefmark{2}$^{\ast}$, Patanamon~Thongtanunam\IEEEauthorrefmark{3}\\
Raula~Gaikovina~Kula\IEEEauthorrefmark{4}, Yasutaka Kamei\IEEEauthorrefmark{2}, Shane McIntosh\IEEEauthorrefmark{1}
}
\IEEEauthorblockA{
\IEEEauthorrefmark{1} University of Waterloo, Canada; \{rungroj.maipradit, shane.mcintosh\}@uwaterloo.ca\\
\IEEEauthorrefmark{2} Kyushu University, Japan; \{d.wang, kamei\}@ait.kyushu-u.ac.jp\\
\IEEEauthorrefmark{3} The University of Melbourne, Australia; patanamon.t@unimelb.edu.au\\
\IEEEauthorrefmark{4} Nara Institute of Science and Technology, Japan; raula-k@is.naist.jp\\
}
\thanks{$^\ast$Corresponding author.}
}
\maketitle
\begin{abstract}
Code review is a popular practice where developers critique each others' changes. 
Since automated builds can identify low-level issues (e.g., syntactic errors, regression bugs), it is not uncommon for software organizations to incorporate automated builds in the code review process.
In such code review deployment scenarios, submitted change sets must be approved for integration by both peer code reviewers and automated build bots.
Since automated builds may produce an unreliable signal of the status of a change set (e.g., due to ``flaky'' or non-deterministic execution behaviour), code review tools, such as Gerrit, allow developers to request a ``recheck'', which repeats the build process without updating the change set.
We conjecture that an unconstrained recheck command will waste time and resources if it is not applied judiciously.
To explore how the recheck command is applied in a practical setting, in this paper, we conduct an empirical study of 66,932 code reviews from the OpenStack community.

We quantitatively analyze (i) how often build failures are rechecked; (ii) the extent to which invoking recheck changes build failure outcomes; and (iii) how much waste is generated by invoking recheck.
We observe that (i) 55\% of code reviews invoke the recheck command after a failing build is reported; (ii) invoking the recheck command only changes the outcome of a failing build in 42\% of the cases; and (iii) invoking the recheck command increases review waiting time by an average of 2,200\% and equates to 187.4 compute years of waste---enough compute resources to compete with the oldest land living animal on earth.

Our observations indicate that the recheck command is frequently used after the builds fail, but does not achieve a high likelihood of build success.
Based on a developer survey and our history-based quantitative findings, we encourage reviewer teams to think twice before rechecking and be considerate of waste.
While recheck currently generates plenty of wasted computational resources and bloats waiting times, it also presents exciting future opportunities for researchers and tool builders to propose solutions that can reduce waste.
\end{abstract}



\section{Introduction}
Code review is broadly recognized as a key approach for improving the quality of software projects~\cite{p42, wang2021can}.
While it was rooted in the formal code inspection process of the past~\cite{s49}, the modern variant of code review that is popular today~\cite{kononenko2016code, mcintosh2016empirical} features online reviewing tools (e.g., Gerrit) and asynchronously performed reviewing activities.
The benefits of code review are well established, including code improvement, knowledge transfer, and finding defects~\cite{Bacchelli_ICSE2013}.

Although code review brings many benefits, it introduces a large overhead for software organizations.
Indeed, Bosu and Carver~\citep{bosu2013impact} estimate that developers spend an average of six hours per week performing reviews.
Recent efforts incorporate automation to reduce the reviewing burden on developers~\cite{panichella2015would, rahman2017impact}, including \emph{automated builds} to relieve the burden of detecting low-level issues (e.g., regression bugs).
Often, automated builds are incorporated through a Continuous Integration (CI) bot, which automatically compiles and tests each version of a proposed change set~\cite{duvall2007continuous}.
The key benefit of CI in this context is to provide quick feedback to developers on whether change sets will smoothly integrate with the existing codebase.
This is especially useful, since both authors and reviewers can focus their discussion on higher-level code properties~\cite{bernardo2018studying, cassee2020silent}, e.g., design~\citep{zanaty2018empirical}.

When the outcome of a CI job is not successful (i.e., a CI job fails), developers may need to diagnose the failure, update their change set, and re-run the CI job. However, it is possible that  the signal from a CI job may not be completely reliable and the diagnosis of a CI failure may suggest that the change set is not to blame~\cite{hilton2017trade}.
For example, a CI job may fail sporadically if tests produce non-deterministic (a.k.a., ``flaky'') results~\citep{luo2014empirical, eck2019understanding, lam2020study}, testing environments encounter infrastructure failures~\citep{9793972}, or a CI service provider has an outage~\citep{gallaba2022lessons}.
When developers suspect that a change set is not to blame for a failing CI job, they may request for the CI bot to repeat the process.
For example, to issue such a request, users of the Gerrit code review tool can post a comment containing the word ``recheck''.

While repeated builds can be justified if the outcome changes, if they are not applied judiciously, additional computing resources can be wasted, and waiting time during the review process can accumulate.
Based on the definition of Sedano et al.~\cite{sedano2017software}, this sort of (unnecessary) computation and waiting time for the repeated CI process with unchanged outcomes is considered \emph{software waste} because the activity consumes resources without providing much value (if any) to the software organization.

Organizational guidelines often attempt to aid in decision-making when it comes to repeated builds.
For example, the testing guidelines for the OpenStack organization\footnote{\label{foot:openstack}\url{https://docs.openstack.org/project-team-guide/testing.html}} state that:
\begin{quote}
    \textit{You will be tempted to just recheck the patch to see if it fails again, but please\textbf{ DO NOT DO THAT}. CI test resources are a very scarce resource (and becoming more so all the time), so please be extremely sparing when asking the system to re-run tests.}
\end{quote}
Although the consumption is often opaque to developers, wasted CI resources consume project finances and can even increase the carbon footprint of an organization~\citep{jiang2014much}.

This paper explores the extent to which developers use the recheck command to repeat the CI automation process during code review, and how much potential waste (waiting time and computational resource) is generated.
We conduct an empirical study of 66,932 code reviews from the OpenStack community.
To structure our study, we formulate the following three research questions:

\begin{enumerate}[\bfseries RQ1.]
    \item \textbf{\RqOneText}\\
\underline{Results.} 55\% of code reviews contain at least one request to repeat the CI process without updating the code via the recheck command. 
On average, recheck is invoked twice within each change set, while 58.5\% of the change sets invoked a single recheck. 
These results suggest that in the OpenStack community, the recheck function is often used to repeat the CI process, but it is typically used sparingly within change sets.

\item \textbf{\RqTwoText}\\
\underline{Results.} We find that the CI outcomes of 42\% of change sets were changed after the recheck commands were invoked, with only 24\% of these attempts being justifiable waste (i.e., invoking recheck only once).
From a closer inspection of the 100 most frequent test jobs that had changed outcomes after rechecks were issued, we find that integration test jobs are the most frequent type of test job associated with changing outcomes. 
These results suggest that CI outcomes do not always change after a recheck is issued. Often, multiple rechecks are needed to obtain a successful CI outcome.

\item \textbf{\RqThreeText}\\
\underline{Results.} We estimate that the total amount of overhead that is generated by the recheck command is a computational time of 187.4 years and an additional review waiting time of 16.81 years, while justifiable waste generates 16.78 years of computational time and 1.66 years of review waiting time.
We also find that reviews where recheck is issued take an average of seven times more computational time and 22 times more waiting time than those reviews without recheck. 
\end{enumerate}

To understand how the OpenStack community perceives our empirical observations, we solicit their input through an online survey.\footnote{
We obtained approval for this study from the Nara Institute of Science and Technology ethics review board on , "A Study of how a Recheck of Software Builds causes Resource Waste in Continuous Integration during Software Development" (\# 2022-I-13).
}
We receive 24 valid responses from community members, with the majority having more than five years of service.
Thematic analysis of the responses suggests that developers frequently invoke recheck despite being aware of the large costs that rechecks generate.

While recheck currently generates plenty of wasted computation and waiting time, our study also presents exciting future opportunities for researchers and tool builders.
Such opportunities include (i) applying restraints, such as timeouts and delays on repeated rechecks, and (ii) proposing mechanisms to recheck only failed jobs.
Much like the key principle in lean manufacturing, we envision opportunities to identify and eliminate such software waste.

\noindent\textbf{Fostering Open Science:} We provide a replication package, which includes raw data, statistical test results, survey results, and a code script~\cite{package}.

\begin{figure*}[]
    \centering
    \subfigure[Change log showing a thread of comments.]{\includegraphics[width=0.55\textwidth]{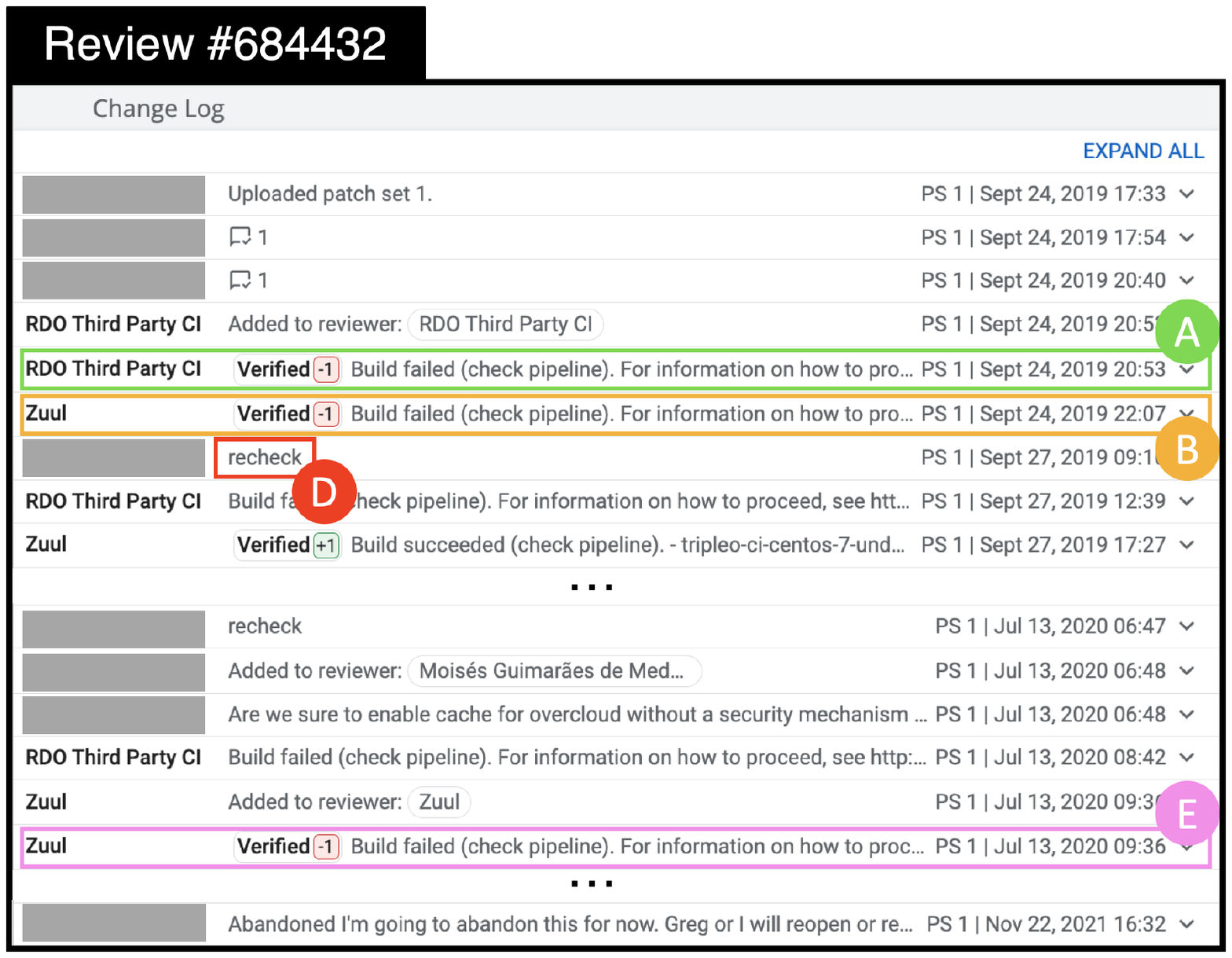}
    \label{fig:motivating_example_a}}
    \subfigure[Test jobs and test outcomes within a CI bot failure.]{\includegraphics[width=0.4\textwidth]{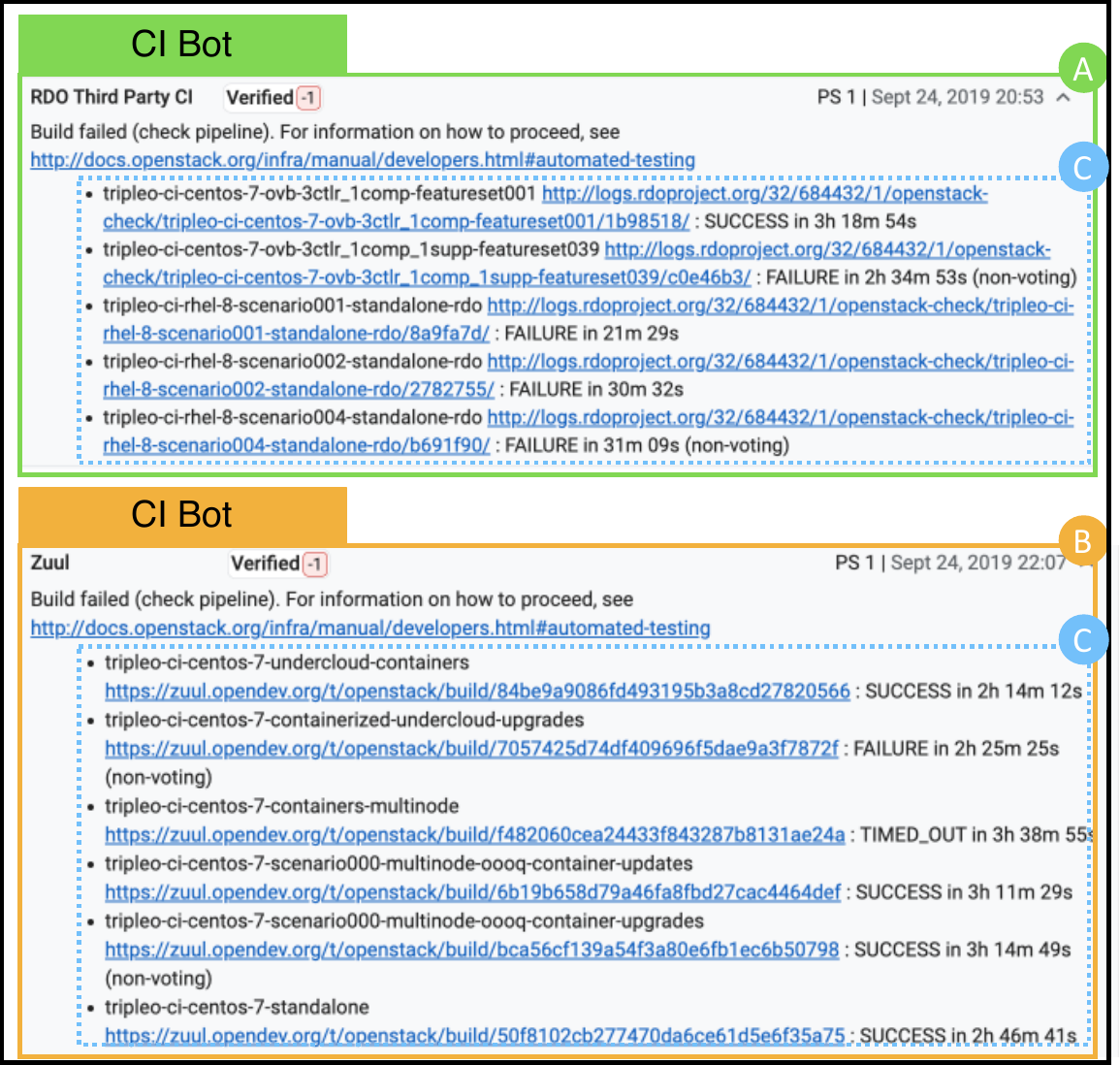}
    \label{fig:motivating_example_b}}
    \caption{Real-world example of the usage of recheck during the code review \#684432 from OpenStack community. \protect\circled{A} box and \protect\circled{B} box denote the CI bot, while \protect\circled{C} box refers to the test jobs in a CI bot. \protect\circled{D} box represents the recheck request.}
    \label{fig:motivating_example}
\end{figure*}

\section{Background and Motivation}
\label{sec:background}
In this section, we introduce the code review process, explain the CI job repetition process, and provide a real-world example of recheck waste to motivate our study.

\textbf{(A) Code Review and Continuous Integration (CI).} 
Modern code review is a lightweight process, which has been adopted in industrial and open-source settings~\cite{FSE2013_Rigby, wang2021can}.
Nowadays, code review is typically performed online and asynchronously using a review tool.
To begin the process, a code author first submits \textbf{a change set}
(i.e., a cohesive set of changes to project files) with a description of the changes to the review tool.
Then, reviewers (i.e., developers other than the author) critique the premise, content, and structure of the submitted change set to provide feedback that the author can use to improve.

A code author may test their change set locally before submitting it; however, not all developers have access to sufficient test resources to run the entire suite of tests.
Hence, reviewers cannot assume that a developer has performed all of the relevant tests. 
Therefore, the code review process is tightly integrated with CI through bots that automatically perform builds and invoke regression tests.
Along with the reviewers' feedback, CI bots report \textbf{the outcome} of the automated build and regression tests.
Figure~\ref{fig:motivating_example} shows an example of a code review ID  \texttt{\#684432} from the OpenStack community.\footnote{https://review.opendev.org/c/openstack/tripleo-heat-templates/+/684432} 
The change log in Figure~\ref{fig:motivating_example_a} shows that the patch set 1 of \texttt{\#684432} received failure reports from two \textbf{build jobs} (i.e., \circled{A} and \circled{B} ) that originate from two CI bots, i.e., \texttt{RDO Third Party CI} and \texttt{Zuul}.

To address reviewer feedback and satisfy CI bots, authors may upload revisions of change sets.
This review, test, and improvement cycle is repeated until the latest revision satisfies the CI bot and the reviewers.
To adhere to Gerrit terminology, we refer to a \textbf{change set} as a \textbf{code review}, and \textbf{revisions of a change set} as a \textbf{patch set}. 

\textbf{(B) Re-run CI job.}
To determine whether and how to address feedback from a CI bot, authors must diagnose log files and perhaps even reproduce the failure locally.
If authors suspect that the failure is not related to their code review, they may request a re-run of the CI bot using the recheck command.
The recheck command is particularly helpful when tests are known to be \textit{flaky}.
For example, Figure \ref{fig:motivating_example_b} shows lists of \textbf{test jobs} and their associated build jobs (see \circled{C}), which indicate that patch set 1 failed due to two test jobs from \texttt{RDO Third Party CI} and one test job from \texttt{Zuul}.
Then, the code author requested a recheck (see \circled{D} in Figure\ref{fig:motivating_example_a}).
Even though multiple rechecks were issued, patch set 1 still failed the regression tests of \texttt{Zuul} (see \circled{E} in Figure\ref{fig:motivating_example_a}).

\textbf{(C) Resource Waste generated by CI Bots.} 
Although the recheck command could save the developer effort on investigating test failures, it is wasteful to recheck without first diagnosing whether the failure is due to flakiness.
When failures are indeed caused by a patch set, blind \texttt{rechecks} waste review waiting time and computational resources and the CI outcome is likely unchanged after a recheck.
Senado et al.~\cite{sedano2017software} argue that any activity that consumes resources but creates no value can be considered as \textit{Software Development Waste}.
Thus, the waiting time and computational resources that are spent on rechecks of non-flaky builds (i.e., CI outcomes do not change) could be considered as software waste as well.

The example in Figure \ref{fig:motivating_example} shows that code author \texttt{\#684432} requested two \texttt{rechecks} for patch set 1 before abandoning the patch set.
As we can see, the outcome of the tests of the \texttt{Zuul} and \texttt{RDO Third Party CI} jobs continued to fail.
The average waiting time for these rechecks is 3 hours 12 minutes.
Moreover, these two rechecks generate an additional 
wasted 6 hours 25 minutes of waiting time, and 41 hours 2 minutes of computational time.
Motivated by this, we set out to examine the prevalence of the use of the recheck command and investigate the potential costs that are incurred.

\section{Case Study Design}
\label{sec:study_design}
In this section, we motivate our research questions and describe our data preparation approach.

\subsection{Research Questions}
Three research questions are formulated to guide our study.

\smallskip
\noindent
\textbf{\RqOne{}}: \textbf{\RqOneText{}}

\smallskip
\noindent
\underline{Motivation.}  Continuous Integration (CI) bots help code authors and reviewers automatically check whether the proposed patch set can be safely merged into an upstream  repository (i.e., pass automated quality checks). If the build is unsuccessful, the code author may use the recheck command to rebuild the patch set without making any code changes.
Since rechecked builds impose additional computational (and in terms, financial) costs, we set out to better understand to what degree CI build failures are rechecked.
    
\smallskip
\noindent
\textbf{\RqTwo{}}: \textbf{\RqTwoText{}}

\smallskip
\noindent
\underline{Motivation.} Rechecks should be used when CI jobs fail due to factors external to the code change under scrutiny (e.g., test flakiness); however, the extent to which these rechecks change the build outcome is unknown. Thus, we formulate RQ2 to investigate the degree to which CI outcomes change after recheck is performed.

\smallskip
\noindent
\textbf{\RqThree{}}: \textbf{\RqThreeText{}}

\smallskip
\noindent
\underline{Motivation.} When recheck is performed,  CI bots spend additional computation resources to rebuild a patch set. In addition, the reviewing process is stalled while waiting for the outcomes of the recheck. 
As discussed in Section 2, the outcome may not change after rechecking, implying that additional computational and waiting time have been spent ineffectively.
Therefore, we set out to quantify this potential overhead.

\subsection{Data Preparation}

\noindent
\textbf{Studied Projects:}
In this paper, we study the OpenStack community because they use the popular Gerrit review tool,\footnote{https://www.gerritcodereview.com/} and the community has been extensively studied in the context of code reviews~\cite{mcintosh2016empirical, wang2021understanding}.
OpenStack is a community of open-source software projects where numerous established organizations and companies collaborate to develop a cloud computing platform. 
The OpenStack community has invested substantial effort in carefully documenting and performing code reviews over the last decade, making it a valuable subject community from which we can draw insights and lessons learned for other communities.
Within the OpenStack community, we have selected the four most active sub-projects: Nova, Cinder, Ironic, and Neutron~\citep{chen2020understanding}.
OpenStack is an open-source cloud computing platform that allows users to manage compute, storage, and networking resources through an API.\footnote{https://www.openstack.org/software/}
These four studied sub-projects, Nova, Cinder, Ironic, and Neutron, represent a diverse range of functionality within OpenStack, which includes virtual machine compute, block storage, bare metal hardware lifecycle, and networking services, respectively.



\begin{table}[]
\centering
\caption{Summary statistic of collected dataset}
\label{tab:data_summary}
\resizebox{.5\textwidth}{!}{
\begin{tabular}{lr}
\toprule
\# Studied Project & OpenStack\\
Studied Period                                & August 2012 -- June 2022       \\
\# Code reviews                             & 66,932      \\
\# Review Comments (change log)            & 898,706 \\
\# Patch sets & 267,239 \\
Avg. - Med. - Max. \# patch sets per CR & 4 - 2 - 948\\
\bottomrule
\end{tabular}}
\end{table}
\textbf{Data collection:}
To collect the code reviews from these four sub-projects, we use the RESTful API provided by the Gerrit code review tool. 
As shown in Table \ref{tab:data_summary}, we capture 66,932 closed code reviews along with 267,239 patch sets from August 2012 to June 2022.
Since we set out to investigate interactions between developers and CI bots, we further collected review comments (which include change log information) related to each code review.
In total, we retrieve 898,706 review comments.

\section{Study Results}
\label{sec:results}
In this section, we present the approach and findings for each proposed research question.

\subsection{Recheck Prevalence}

\begin{figure}
    \centering
    \includegraphics[width=.9\linewidth]{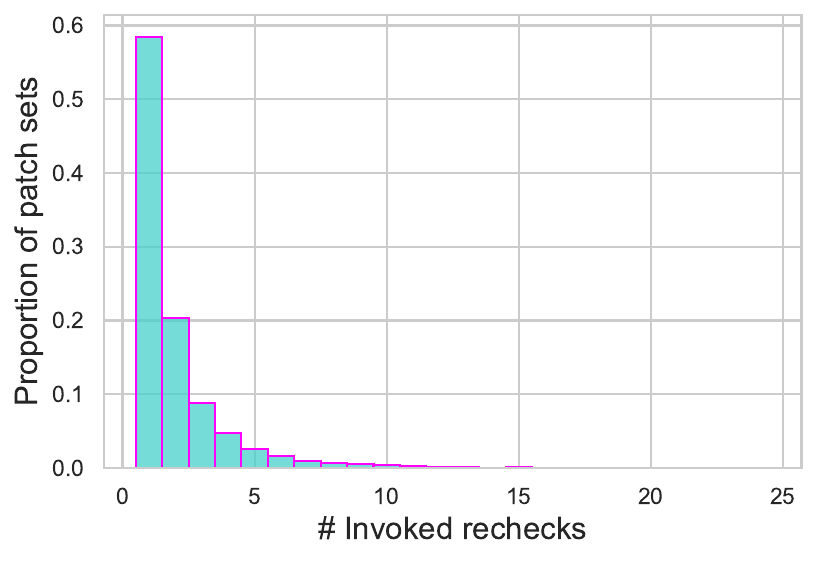}
    \caption{Proportion of patch sets and number of invoked rechecks.} 
    \label{fig:patch_set}
\end{figure}

\textbf{\underline{Approach.}}
To investigate to what degree are CI build failures rechecked (\RqOne{}), we  analyze descriptive statistics of the number of patch sets that (i) did not pass CI bots (i.e., received build failures from CI bots) and (ii) were rechecked.
Specifically, to identify patch sets with build failures, we use a regular expression, i.e., \texttt{(?:build failed)} to search the build failure outcomes in comments from CI bots in the studied code reviews.
Then, we check if recheck is requested for these patch sets that received build failures from CI bots using a regular expression, i.e., \texttt{(?:recheck)} on the comments posted after the build failures for these patch sets.

To quantify the prevalence of CI builds that are rechecked, we first calculate the proportion of code reviews that have at least one patch set with build failures and those build failures were rechecked.
Since one code review may have multiple patch sets, we also calculate the proportion of patch sets that were rechecked after the build failures.
It is also possible that one patch set may have multiple rechecks.
Thus, we further examine the frequency and distributions of rechecks that were performed per patch set.

\begin{figure*}
    \centering
    \includegraphics[width=0.8\textwidth]{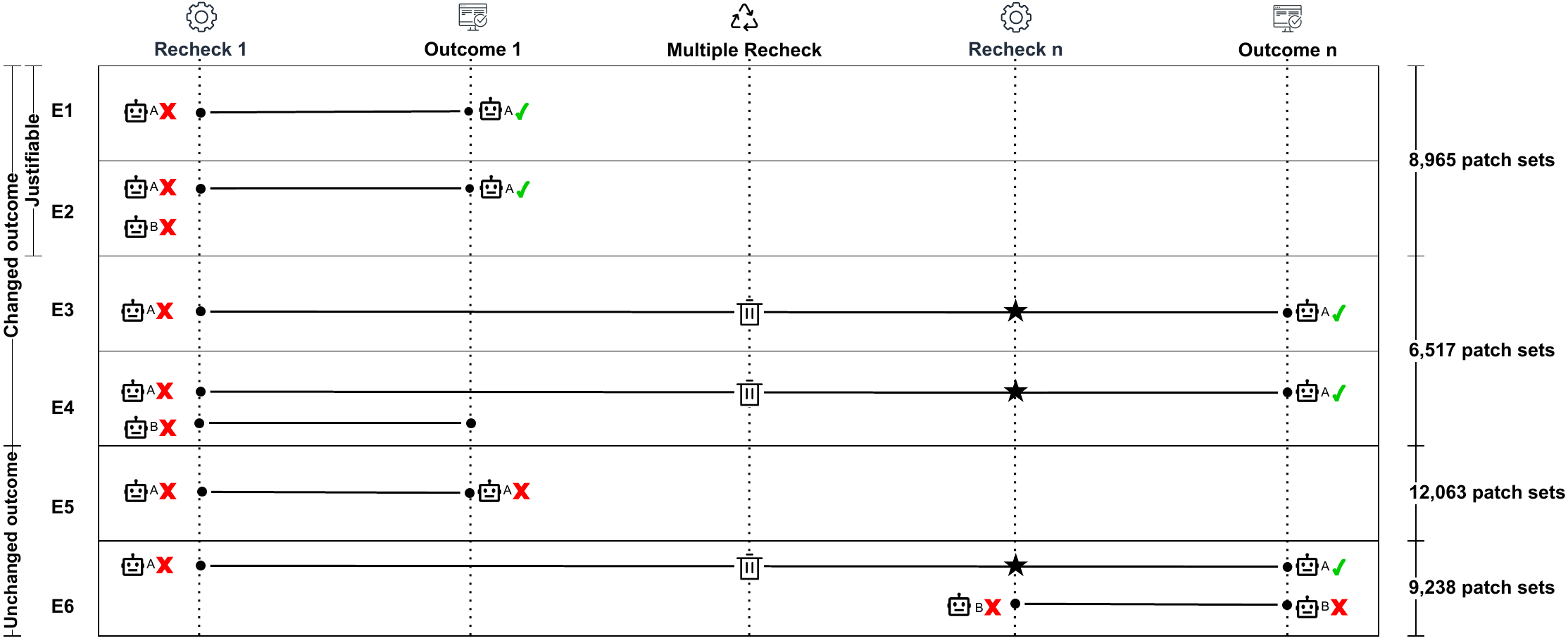}
    \caption{Six examples to illustrate possible outcomes of CI failed jobs with an invoked recheck to repeated the build.
    }
    \label{fig:my_label}
\end{figure*}

\smallskip
\noindent
\hypertarget{result:obs1}{\textbf{Observation 1:}} \ul{\textbf{55\% of code reviews have at least one patch set that was rechecked after the failing CI job.}}
From 41,868 code reviews that contain at least one build failure, we find that more than half of the studied code reviews (55\%; 23,086 out of 41,868 reviews) with at least one patch set that was rechecked after the build failures. 
As shown, 45\% of code reviews did not invoke rechecks during the code review process. 
In this case, the test jobs had failed without recheck invoked.
Investigations at the patch set granularity level indicate that a bit over a quarter of the patch sets (i.e., 26\%; 36,783 out of 142,165) that had a recheck invoked after CI failed, which is a one in four chance to invoke a recheck.   

We also investigated the prevalence of recheck commands in passing builds to further our understanding of the recheck commands used to repeat the CI process. Among 80,428 patch sets containing only build success, we discovered that 2,307 (3\%) contained recheck commands.
One reason for using recheck in passing builds seems to be related to failing tests with non-voting power (i.e., tests whose result does not determine whether the build is a success or failure), with 55\% of all passing builds containing recheck including such tests.
For example in the code review ID \#752206\footnote{\url{https://review.opendev.org/c/openstack/tripleo-ci/+/752206/1}} first patch set, after developers attempt to recheck, a comment mentions an error about failing non-voting jobs.

\smallskip
\noindent
\hypertarget{result:obs2}{\textbf{Observation 2:}} \ul{\textbf{On average, a patch set with build failures was invoked a recheck two times.}}
Figure~\ref{fig:patch_set} shows the distribution of invoked rechecks that occur in a patch set.
In detail, the figure shows the proportion of patch sets (y-axis) against the number of invoked rechecks.
According to the figure,  most patch sets (i.e., 59\%) invoked a single recheck after a build failure.
In the extreme case, the number of recheck in a patch set could reach up to 226. 
We observe that in this case, the code review is often submitted to test CI jobs.
An average mean, a patch set is prone to multiple rechecks (having two times) after the CI build failure.

\begin{tcolorbox}\textbf{RQ1 Summary:}
Findings indicate that more than half (i.e., 55\%) of code reviews with failing CI jobs were rechecked by the review team. 
Furthermore, most patch sets (59\%) attempted a single recheck.
\end{tcolorbox}



\subsection{Changing CI Build Outcomes}

\textbf{\underline{Approach.}}
To investigate the extent to which invoking recheck changes the CI outcome from failure to success (RQ2), we analyze (i) the proportion of the patch sets whose CI outcomes were changed after invoking recheck and (ii) the kinds of test jobs in the changed CI outcomes after rechecks.
We analyze the 36,783 patch sets that had at least one recheck invoked after failing CI jobs.

\noindent
\textbf{Proportion of changed CI Outcomes:}
To investigate this, we first extract the last CI outcomes of patch sets after rechecks. 
To do so, we apply the regular expression, i.e., \texttt{(?:success | failure| aborted)} on the CI reports in the change log (i.e., comments from CI bots).
Then, we determine patch sets as having \textit{changed} CI outcome after rechecks if their last CI outcomes after recheck are successful.
In addition to the proportion of patch sets with changed CI outcomes, we also examine the number of rechecks that were invoked until the last CI outcomes.

Figure \ref{fig:my_label} illustrates different examples showing different scenarios of CI outcome changes and whether rechecks are justifiable.
Note that a build of a patch set may contain one or more bots. 
Based on the review, CI bots are either automatically or manually added, removed, or replaced for a build run.
For the first four examples (E1 - E4), we consider these patch sets as having \textit{changed} CI outcomes after rechecks because the last CI outcomes are successful. 
Note that in the second example (E2), recheck was not invoked for CI bot \texttt{B}, thus we do not consider CI bot \texttt{B} in this scenario.
If the CI outcome changed after a single recheck, we consider this recheck as \textit{justifiable}, i.e., an initial CI job may fail due to CI-related issues (e.g., flaky tests, test environments), not the code in the patch set.
In Figure \ref{fig:my_label}, we determine the rechecks in the first two examples (E1 and E2) as \textit{justifiable}, while the rechecks in the other examples are considered as \textit{non-justifiable}.
Although the CI outcome changed after multiple rechecks (E3 and E4), we argue that such rechecks could be avoided if the developers figure out the causes, hence we regard these cases as non-justifiable.
On the other hand, if at least one of the last CI outcomes is still a failure, we consider these patch sets as having \textit{unchanged} CI outcomes after rechecks.
For example, we consider the patch sets in the last two examples (E5 and E6) in Figure \ref{fig:my_label} as having \textit{unchanged} CI outcomes after rechecks.

\noindent
\textbf{Test Jobs in Changed CI Outcomes:}
To better understand the CI jobs that have changed outcomes, we manually investigate the kind of test jobs that have changed the outcome from failure in the initial CI job to success in the last CI jobs.
We first extract the test jobs and their outcomes in the initial CI outcomes and the last CI outcomes for each patch set using the regular expression, i.e., \texttt{((?:success | failure | aborted) in(?: \textbackslash w+?)*s)}.
Figure \ref{fig:motivating_example_b} \circled{C} shows an example of test jobs and their outcomes that we extracted.
Then, we identify the test jobs whose outcomes in the initial CI jobs were changed from failure to success in the last CI outcomes as test jobs with \textit{changed outcomes}.
Note that we examine the initial and last CI outcomes that were from the same CI bot.
In total, we have identified 1,765 test jobs with the changed outcome which are from 4,098 distinct test jobs. 

Since the 100 most frequent test jobs whose outcomes changed after rechecks account for 51\% of all test jobs with changed outcomes, we manually examine the kinds of these 100 most frequent test jobs.
For the kinds of test jobs, we refer to the OpenStack official documentation related to CI\footnote{\url{https://docs.openstack.org/project-team-guide/testing.html}} which identifies five main kinds of test jobs: (I) \textit{integration test}, (II) \textit{unit test}, (III) \textit{functional test}, (IV) \textit{upgrade test}, and (V) \textit{style checks}.
The first two authors independently classify each of the 100 most frequent test jobs with changed outcomes based on the CI reports, the OpenStack job documentations\footnote{\url{https://zuul.opendev.org/t/openstack/jobs}} and information from GitHub mirror repositories.\footnote{\url{https://github.com/openstack}}
The inter-rater agreement (cohen's kappa) between the first two authors are 0.78 (i.e., substantial agreement).
Finally, the first two authors and the fourth author discussed and resolve the disagreement on the kinds of 12 studied test jobs.

\begin{table}[]
\centering
\caption{Proportion of patch Sets with changed and \\ unchanged CI outcomes after rechecks}
\label{tab:rq2_pattern}
\begin{tabular}{llrr} 
\toprule
\multicolumn{2}{l}{CI Outcome} &  Count & Percentage \\ 
\hline
\multirow{3}{*}{\begin{minipage}{2.5cm}Changed outcome \newline (Failure $\xrightarrow{}$ Success)\end{minipage}} & Justifiable recheck & 8,965 & 24.37\% \\ 
 & Multiple rechecks & 6,517 & 17.72\% \\ \cline{2-4}
& Total & 15,482 & 42.09\% \\ 
\hline
\multirow{3}{*}{\begin{minipage}{2.5cm}Unchanged outcome \newline (Failure $\xrightarrow{}$ Failure)\end{minipage}} & Single recheck & 12,063 & 32.80\% \\ 
 & Multiple rechecks & 9,238 & 25.11\% \\ \cline{2-4}
  & Total & 21,301 & 57.91\% \\
\bottomrule
\end{tabular}
\end{table}

\smallskip
\noindent
\hypertarget{result:obs3}{\textbf{Observation 3:}} \ul{\textbf{Of all failing CI jobs that are rechecked, only 42\% of them change the build outcome.}}
Table \ref{tab:rq2_pattern} presents the results of build outcome change after the recheck.
On the one hand, we observe that, less than half of build failures (42\%; 15,482 out of 36,783) changed the outcome (i.e., from failure to success) after invoking the recheck command on the patch set.
Furthermore, 24\% of build failures are justifiable, whereas 17\%  of them required multiple rechecks to eventually get the CI outcome to change.
On the other hand, our results show that a relatively large proportion (33\%) of patches with a single recheck did not receive changed CI outcomes, suggesting that rechecks are less likely to change CI outcomes.

Additionally, we conduct an analysis on the reassignment of CI bots to each patch set to gain a better understanding of how different CI bots change the outcome. The data reveal that for both justifiable repeat builds and successful patch sets with multiple rechecks, there are more likely to assign the same bots
(i.e., CI bots are the same before and after invoking recheck) at rates of 66\% and 57\% respectively. This is different from the unchanged outcome for single rechecks and multiple rechecks which were 37\% and 45\% respectively.
Further investigation showed that the \texttt{Zuul} CI bot was prevalent in repeated CI builds that eventually changed outcomes, with a success rate of 63\% (running 21,084 times).


\smallskip
\noindent
\hypertarget{result:obs4}{\textbf{Observation 4:}} \ul{\textbf{Integration test jobs are more likely to change the build outcomes after rechecks.}}
As we can see, the integration test job (i.e., designed to validate the project and all related component) tends to change the outcomes, when compared with other test jobs, accounting for 60\%.
The second most common test job is the functional test (17\%), which is designed to validate requirements and verify that the output is consistent with user expectations.
While the rest of test jobs are unit test, upgrade test, and style check at (11\%, 11\%, and 1\%) respectively.
This observation also helps to explain why there are so many build failures. 
One of the possible reasons is the large number of integration tests which have a high chance of producing a faulty build.

\begin{tcolorbox}\textbf{RQ2 Summary:}
When invoking a recheck, we find that less than half (i.e., 42\%) are able to successfully change the build outcomes.
Furthermore, only 24\% were justifiable waste (i.e., having a single recheck to change the outcome).
Qualitatively, we find that rechecks on integration test jobs are more likely to change the build outcome.
\end{tcolorbox}

\subsection{Overhead generated by rechecks}
\textbf{\underline{Approach.}}
To address RQ3, we quantify software waste that is generated by invoking recheck after the CI build failures.
In this RQ, we analyzed the 36,783 patch sets where at least one recheck was invoked.
Based on prior work~\cite{sedano2017software}, we examine software waste in terms of \textit{waiting time} and \textit{computational time}.
Due to the parallel nature of the CI build process, we expect build times to exceed actual review times, as we account for overlaps in the process.
Hence, our metrics should quantify the extent of this overlap.
We now describe how we measure the waiting time and computational time.

\noindent
\textbf{Quantifying Waste:} Sedano et al. identified waiting/multitasking/parallel due to slow tests or unreliable tests or unreliable acceptance environment as a form of software wastes~\cite{sedano2017software}.
Align with this definition, in this context, the \emph{\textbf{waiting time}} for new CI outcomes after rechecks can be considered as software waste.
Therefore, we measure the waiting time by identifying the longest time that CI jobs took to complete a recheck (i.e., repeated builds and tests). 
Since one CI job can have multiple test jobs which can be executed concurrently, we consider the longest time that the test jobs spent.


In a lean manufacturing context (i.e., manufacturing waste), extra processing time can be considered as waste~\cite{sedano2017software}.
Thus, we are also interested to quantify the \emph{\textbf{computational time}}, i.e., in the computational time that CI servers spent to repeat the build and test jobs. 
We measure the computational time by accumulating the running time of all test jobs of CI bots in all rechecks that were invoked for a patch set.

\begin{figure*}[]
    \centering
    \subfigure[Review time for a patch set (log hours)]{\includegraphics[width=0.45\textwidth]{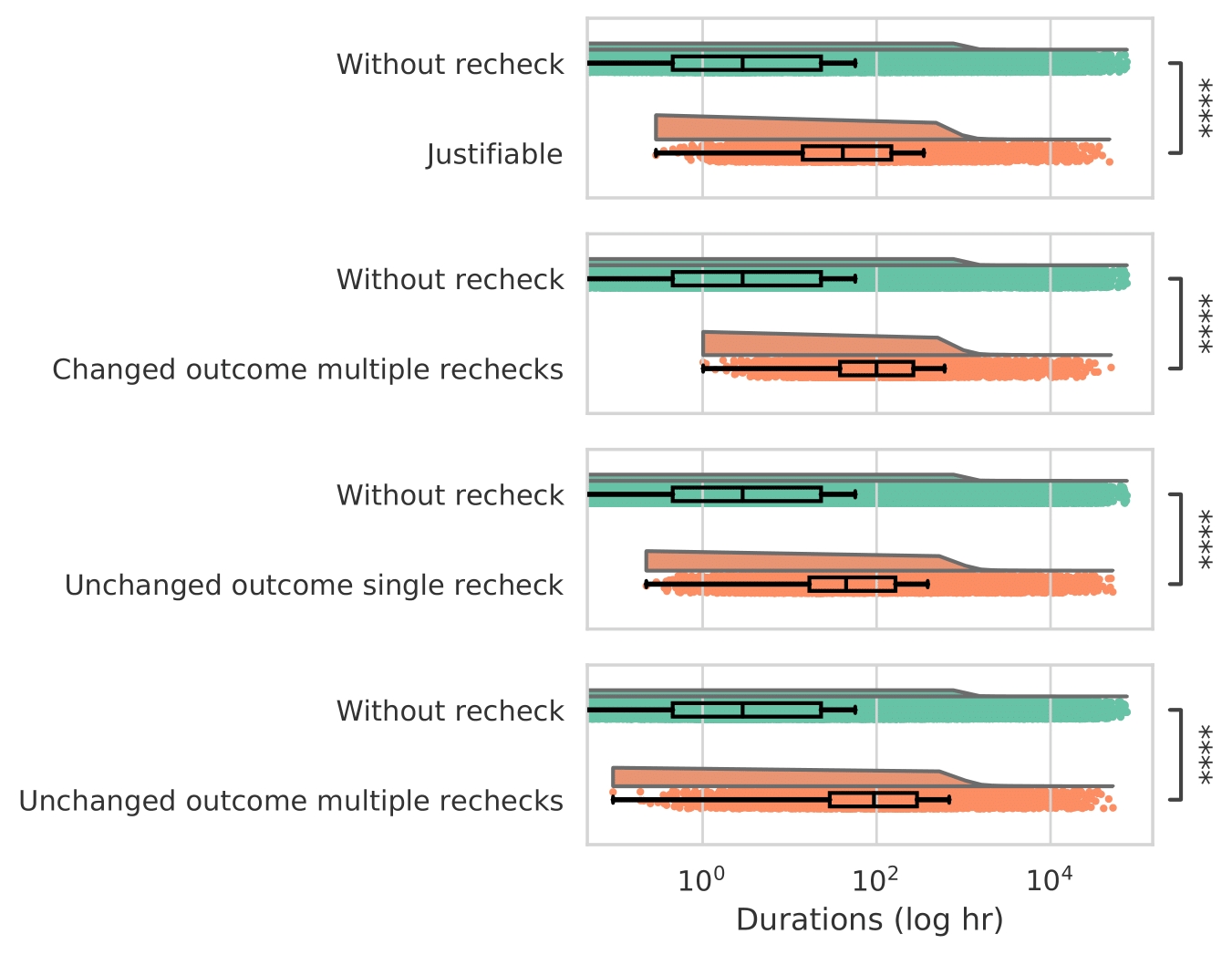}
    \label{fig:RQ3_review_time}}
    \subfigure[Computational time for a patch set (log hours)]{\includegraphics[width=0.45\textwidth]{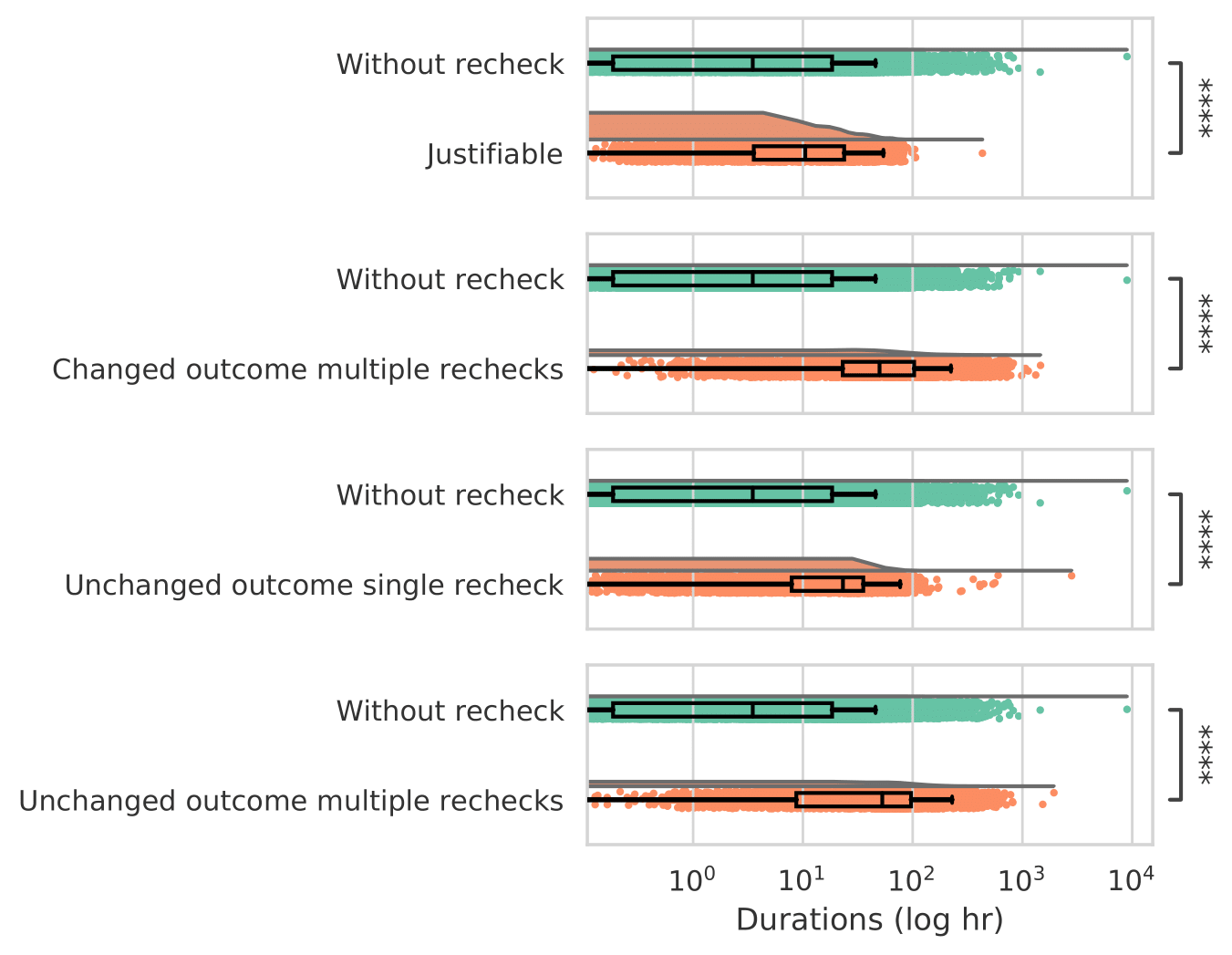}
    \label{fig:RQ3_computation_time}}
    \caption{Comparison of (a) review time (b) computational time between with and without recheck (Mann-Whitney U test ****: p <= 0.0001)}
    \label{fig:RQ3}
\end{figure*}

\begin{table}[]
\centering
\caption{Accumulative waiting time generated by recheck}
\label{tab:waiting_cost}
\begin{tabular}{llcr}
\toprule
 Outcome & & Time (Median) & Accumulated \\
\midrule
\multirow{2}{*}{Changed} & Justifiable & 1hr 34min & 1.66 years \\ 
& Multiple rechecks & 5hr 7min & 5.18 years\\ \midrule
\multirow{2}{*}{Unchanged} & Single recheck & 2hr 22min & 3.32 years \\ 
& Multiple rechecks & 5hr 9min & 6.65 years\\
\midrule
Total & & 2hr 35min & 16.81 years \\
\bottomrule
\end{tabular}
\end{table}

\begin{table}[]
\centering
\caption{Accumulative computational time generated by recheck}
\label{tab:computation_cost}
\begin{tabular}{llcr}
\toprule
 Outcome & & Time (Median) & Accumulated \\
\midrule
\multirow{2}{*}{Changed} & Justifiable & 10hr 30min & 16.78 years \\ 
& Multiple rechecks & 49hr 54min & 62.57 years\\ \midrule
\multirow{2}{*}{Unchanged} & Single recheck & 23hr 10min & 34.62 years \\ 
& Multiple rechecks & 52hr 51min & 73.43 years\\
\midrule
Total & & 25hr 15min & 187.40 years \\
\bottomrule
\end{tabular}
\end{table}
\noindent
\textbf{Analyzing waste:} To do so, we compute an average waste (i.e., waiting time and computational time) per patch set and the sum of total waste that was generated over the studied period (i.e., 10 years).
We also analyze the waste in the patch sets that have recheck and patch sets without recheck.
Patch sets without recheck are identified using a regular expression, i.e., \texttt{(?:recheck)} to search for recheck requests. We select all patch sets that do not include recheck requests.
The analysis will signal the impact of recheck on the efficiency of code reviews compared to regular patch sets.


To make a comparison against patch sets that did not invoke a recheck, we defined two metrics. 
This first metric is the review time, which refers to the duration from when the patch set is uploaded to the last activity of this patch set.
The second metric is the computational time.
For patch sets that did not invoke a recheck, we measure the total CI bot running time. 
For patch sets that repeat the CI build (with recheck), we only consider the CI bots that repeat CI build (after the initial failure). 
We use the Mann-Whitney U test (a non-parametric test)~\citep{mann:1947} and the effect size using Cliff's $\delta$ (a non-parametric effect size measure)~\citep{Cliff:1993} to quantify the difference.

\smallskip
\noindent
\hypertarget{result:obs5}{\textbf{Observation 5:}} \ul{\textbf{Around 16.81 years were spent waiting around for rechecked CI bots to complete their repeated builds}}
Table \ref{tab:waiting_cost} shows waste in terms of the waiting time for the CI bot to repeat the build over a 10-year period (i.e., August 2012 to June 2022).
In total, the review project accumulated 16.81 years of waiting of which 15.15 years were spent on unjustifiable waste.
Breaking down this table, justifiable waste was the lowest, with waiting times accumulating to 1.66 years, while a total of 5.18 years were spent waiting patch sets with multiple rechecks to complete the job, with 3.73 years (72\%) spent on an attempted recheck that produced an unchanged CI outcome and 1.45 years (28\%) spent on the last recheck that produced a changed CI outcome.
On the other hand, Table \ref{tab:waiting_cost} also reports a total of 9.97 years spent on waiting for rechecks that did not change the outcome. 
In detail, failed rebuilds after a single recheck accumulated in 3.32 years spent, while review teams spent a total of 6.65 years waiting on builds with multiple rechecks.

\smallskip
\noindent
\hypertarget{result:obs6}{\textbf{Observation 6:}} \ul{\textbf{A total of around 187 years (computational time) were spent on CI Bots running repeated builds.}}
Similar to Observation 5, Table \ref{tab:computation_cost} reports resource waste in terms of the CI bot computational time. 
In total, 187.40 years of CI bot run-time was spent on repeating builds with 170.62 years spent on unjustifiable.
Breaking this down, only 16.78 years are for justifiable and 62.57 years are for builds with multiple rechecks, with 43.54 years (70\%) spent on an attempted recheck that produced an unchanged CI outcome and 19.03 years (30\%) spent on the last recheck that produced a changed CI outcome.
In terms of failed builds, an accumulation of 108.05 years was spent without any change to the outcome. 
Breaking this down, failed single rechecks wasted 34.62 years, while a repeated build with multiple rechecks consumed 73.43 years of CI bot running time.

Additionally, we estimate the financial impact of generating waste on additional service using circleCI, a dedicated cloud-based CI provider, that charges between 5 and 2,000 credits per minute for customized resources,\footnote{\url{https://circleci.com/product/features/resource-classes/}} with the cost of 1 credit approximately 0.0006 USD.\footnote{\url{https://discuss.circleci.com/t/performance-plan-billing-update/39502}}
Considering the minimum requirement for deploying Murano, one of the systems in OpenStack requires 8 GB of RAM and a CPU with four cores that costs at least 20 credits per minute.\footnote{\url{https://docs.openstack.org/murano/rocky/admin/deploy_murano/prerequisites.html}}
During the 10-year lifetime of OpenStack, it generated 170.62 years of unjustifiable waste, which amounted to 89,739,295 minutes.
In total, the estimated cost for rechecking is 1,076,871.54 USD.

\smallskip
\noindent
\hypertarget{result:obs7}{\textbf{Observation 7:}} \ul{\textbf{
On average, patch sets with repeated builds take considerably more time (2,200\%) to be reviewed compared to patch sets that do not invoke a recheck.
}}
Figure \ref{fig:RQ3_review_time} shows that the patch sets that contain repeated builds tend to have a longer reviewing time.
Table \ref{tab:rq3-stat} also confirms statistical significance that a repeated rebuild will take longer to review.
As these results are intuitive (repeat builds always take longer), we compare the average (median) patch set to compare the extent of time taken for reviews with or without a repeated build.
First, we find that on average, it takes around 2 hours 53 minutes to complete a review without any repeated builds.
Meanwhile, justifiable repeat builds took 40 hours 56 minutes to review, which less time compared to reviews with multiple rechecks (i.e., taking 99 hours 49 minutes). 
In terms of failed builds, a single recheck amounted to 44 hours 38 minutes on average for a review, while repeated failed rebuilds containing multiple rechecks took up 93 hours 15 minutes of reviewing time.

Furthermore, we estimated the potential time savings that could be achieved by avoiding unjustifiable rechecks. We compared the overall time spent on code review with the time spent after the recheck is invoked and found that on average (median), avoiding unjustifiable rechecks could save 92\% of the review time.

\begin{table}[]
\centering
\caption{Statistical comparisons of waste between with and without recheck}
\label{tab:rq3-stat}
\begin{tabular}{@{}lcc@{}}
\toprule
\textbf{Without check} & \textbf{Review time} & \textbf{Comp time} \\ \midrule
\textless~Justifiable & Large & Small \\
\textless~Changed outcome multiple rechecks & Large & Large \\
\textless~Unchanged outcome single recheck & Large & Large \\
\textless~Unchanged outcome multiple rechecks & Large & Large \\
\bottomrule
\multicolumn{3}{l}{\small{Effect size: Negligible $|\delta| < 0.147$, Small $0.147 \leq |\delta| <0.33$,}}\\
\multicolumn{3}{l}{\small{Medium $0.33 \leq |\delta| <0.474$, Large $0.474 \leq | \delta|$}}\\
\end{tabular}
\end{table}

\smallskip
\noindent
\hypertarget{result:obs8}{\textbf{Observation 8:}} \ul{\textbf{
On average, patch sets with repeated builds consumes more (700\%) computational time when compared to patch sets without recheck.
}}
Figure \ref{fig:RQ3_computation_time} is also intuitive, with the CI bot running for reviews with repeated builds longer that those without repeat builds.
Statistical significance is also confirmed in Table \ref{tab:rq3-stat}.
In terms of computational time, a patch set without a repeat build takes on average (median) 3 hours 29 minutes for the CI bots to run.
Justifiable CI bot run times amounted to a total of 34 hours 30 minutes, which was the least amount of waste generated by the repeated builds on average, when compared to failed repeated CI builds (single recheck is 95 hours 10 minutes and multiple rechecks is 52 hours 51 minutes).

Similar to Observation 7, we estimated the potential computation time saving that could be achieved by avoiding unjustifiable rechecks. We compared the total compute time spent on a patch set with the compute time spent after the recheck is invoked and found that and found that on average (median), avoiding unjustifiable rechecks could save 63\% of the compute time.

\begin{tcolorbox}\textbf{RQ3 Summary:}
Overall, the total accumulated waste generated by rechecks is a computational time of 187.4 years, with a waiting time of 16.81 years.
We also find that on average, it takes seven times more computational time and 22 times more waiting time compared to reviews without rechecks. 
\end{tcolorbox}

\section{Community Perception}
\label{sec:implication}


To better understand how developer perceive the usage of the recheck function, we conduct a survey of OpenStack developers, striving to (i) understand how they handle test failures,
(ii) solicit developer feedback about our observations and (iii) our recommendations for future work.
We sent our online survey invitation to 936 developers who have invoked the recheck function at least one time in the past three years based on our studied dataset.
The survey was open from September 29 to November 14, 2022.
We received valid responses from 24 developers.
16 of the 24 respondents have more than five years of contributing to OpenStack projects and 20 of the 24 contribute as both contributors and reviewers.
To analyze the responses of the open-ended questions, we use the card sorting method.
Below, we present our survey questions and discuss the survey results.
The survey design and its sanitized responses are available in our online appendix~\cite{package}.

\textbf{Handling test failures.} In the first portion of our survey, we ask respondents about how they handle test failures.
\ul{Related to the frequency of using recheck}, the majority of respondents report that they (58\%) frequently invoke recheck, i.e., 50\% or 75\% of the times when CI jobs fail.
\ul{For the awareness of official guidelines}, only five respondents admit that they were unaware of the guidelines.
When asked about the practicality of the guidelines, all respondents agree that it is practical to examine the logs of the jobs that failed.
20 of the 24 respondents also agree that it is practical to try whenever possible to reproduce the failure locally.
In the open-coded responses of freeform comments, seven respondents pointed out impractical scenarios for locally reproducing test failures.
For example, one respondent stated that the advice to \textit{``Reproduce locally is OK when the failure is in functional tests, but not for the tempest-based CI jobs''.}
The guidelines also suggest reaching out to the project team, opening a bug report to track the inconsistent test behaviour, and referencing that bug when invoking recheck. Only 9\% and 6\% agree.
Five respondents expressed concerns about the pace at which these bug reports will generate a reaction (if at all). 
For example, one respondent stated \textit{``Issuing the bug doesn't guarantee any feedback from the team.''} and another respondent similarly reported \textit{``Bugs opened for CI failures (exception: across the board CI breakage) are unlikely to be followed up on and even if they are acted upon, it will usually take days to weeks''}.

\ul{In terms of the personal workflow}, 13 of the 24 respondents claimed that they have their own workflow to handle failures.
Seven of them further shared their experience in freeform comments. 
Their coded responses indicate that all of them converge towards the official guidelines. 
For instance, a respondent explained that they \textit{``examine logs, try to find a reason why it failed. If possible, reproduce locally, come up with a fix''}.
\ul{When asked about oversight}, only 6 of the 24 respondents reported that there is oversight in the usage of recheck, such as peer pressure from reviewers and the technical committee, which can condemn projects for not adhering to policy.

\textbf{Feedback on our findings.} In the second portion of our survey, we selected six of our observations [1,2,3,4,7,8] and ask respondents whether their personal experience aligns with them.

In terms of \hyperlink{result:obs1}{Observation 1}, 20 of the 24 respondents stated that they are aware of this situation and have experience.
Eight respondents pointed out that over-reliance on recheck is due to external factors, such as dependencies and unreliable resources.
One respondent highlighted that \textit{``I believe there are people who do a ``blind recheck'' which is bad because it is wasteful''}.

In terms of \hyperlink{result:obs2}{Observation 2}, we similarly find that 20 of the 24 respondents are aware that patch sets with build failures invoked recheck twice on average and report that the observation agree with their personal experience.
Several respondents answered that such a finding is not surprising to them.  
For instance, one of them cited \textit{``So on average people recheck twice. Interesting. Sounds about right''}.
Another respondent believed that rechecking twice can be avoided and suggested that in such cases, developers should create a bug report and wait until it is resolved before invoking a (second) recheck.

With respect \hyperlink{result:obs3}{Observation 3}, 13 of the 24 respondents state that they are aware of this and have similar experiences.
Two respondents felt that the likelihood of the outcome remaining unchanged is higher in their experience.

Regarding \hyperlink{result:obs4}{Observation 4}, 17 of the 24 respondents are aware that integration test jobs are more likely to change build outcomes after rechecking.
Moreover, eleven respondents explained their rationales.
They believe that the unit and functional tests are highly stable and can be reproduced locally; 
however, integration tests contain complex dependencies and are likely to be affected by a bug in one of the multiple projects that are coordinating.
Meanwhile, one respondent provided a feature suggestion for the existing system, stating that they \textit{``100\% agree. That's why I think if a patch set doesn't change and recheck is typed, only recheck the red jobs. No need to recheck the green jobs''}.

Regarding \hyperlink{result:obs7}{Observation 7} and \hyperlink{result:obs8}{Observation 8}, 17 of the 24 respondents state that they are aware of the massive cost of the computation and review time resulting from the recheck builds.
Specifically, five respondents showed their surprise and interest in our observations.
One respondent cited \textit{``I'm surprised the additional compute time is so high, I would have thought 400\% max''}.
Meanwhile, another respondent pointed out a chilling effect that unscrupulous rechecks may have on reviewers, stating that: \textit{``As a core reviewer, if I see changes with a lot of recheck I am less likely to review it right away''}. 

\textbf{Perception of our recommendations.} In the last portion of our survey, we asked respondents to weigh in on our recommendations for code authors, reviewer teams and the software engineering community as a whole:
 \begin{itemize}
    \item \textit{Think twice before rechecking}, our results indicate that only 24\% of repeated builds belong to justifiable waste (\hyperlink{result:obs3}{Observation 3}), so maybe you have only a one in four chance of proving the bot wrong.
    \item \textit{Try to use recheck judiciously}, as our results show that 42\% of rechecks change the build outcome (\hyperlink{result:obs3}{Observation 3}). 
    Even with multiple recheck attempts, only 18\% succeed in changing the outcome.
    \item \textit{If inevitable, only attempt rechecks on Integration tests (\hyperlink{result:obs4}{Observation 4})}, our results indicate that integration tests are far more likely to change outcomes when rechecking.
    \item \textit{Be considerate of waste}, although these costs are hidden, developers should be aware of the computation and waiting time (\hyperlink{result:obs5}{Observation 5, 6, 7, 8}).
\end{itemize}
Based on their responses, we find that 19, 23, 18, and 20 of the 24 respondents agree that the four provided suggestions are practical, respectively.

We also ask respondents about challenges and directions for future work. Respondents ranked \textit{``understanding the reasoning behind a recheck''} and \textit{''making the review team aware of waste``} as the most promising.
In freeform comments, seven respondents shared a common suggestion about a future feature where recheck is constrained to only re-execute failed jobs.
Meanwhile, another respondent suggested that \textit{``I think making the developer write a compulsory explanation for running the recheck seems reasonable''}.


\section{Threats to Validity}
\label{sec:threats}
We breakdown the threats into three parts, external, construct, and internal validity. 

\subsection{External Validity}
These threats relate to the ability to generalize based on our results.
In this study, we conducted an empirical study only focusing on the OpenStack community.
The observations may not be generalized to other open source communities and other technologies other than Gerrit.
While, we are confident that OpenStack is representative enough to be studied, as it actively performs code reviews and is widely considered in code review research (cf. Section VII).
In future work, we would like to extend our framework to other popular open-source communities, such as Qt.
Nevertheless, to encourage research replication, we prepare an online replication package~\cite{package}.
Another threat lies in generalizing beyond the specific context of the study due to the exclusion of code change factors and collaborative aspects of code review. As a result, the findings may not be fully applicable to code review scenarios where these omitted factors play significant roles

\subsection{Construct Validity}

These threats relate to the degree to which our measurements capture  the relevant information. 
To identify the build failures and recheck commands in a code review, we rely on the regular expression to match the review comments.
False positives may occur due to the regular expression design, however, we manually validated the samples of the retrieved information and were sure about its accuracy.
Another threat is concerning the metrics that we use to measure waste, especially when we measure the accumulation of parallel patch sets being executed.
Sedano et al.~\cite{sedano2017software} argue that identifying waste is difficult, as human cognition and status quo bias may hinder practitioners’ ability to notice waste. 
We also claim that as an initial study, with comparisons just to provide context, the first step is to identify waste, with follow-up studies aiming to later redefine these metrics for more precision.

\subsection{Internal Validity}
These threats refer to the approximate truth about inferences regarding cause-effect or causal relationships.
The first threat occurs during our manual analysis of test jobs in RQ2, where the codes may be mislabeled.
To relieve this threat, an open discussion was conducted between two authors and the fourth author joined the discussion to resolve each disagreement. 
The second threat exists in the selection of statistical significance testing techniques in RQ3.
To measure the significance and effect size, we leverage the Mann-Whitney U test and Cliff's $\delta$.
However, we are confident that these testing techniques widely used in the prior work are an appropriate choice.

\section {Related Work}
\label{sec:related}

\textbf{Code Review.}
Nowadays, contemporary software development teams have widely adopted tool-based code reviews for their software quality assurance~\cite{Bacchelli_ICSE2013,FSE2013_Rigby, wang2021can}.
A large body of studies has investigated code review from technical and social perspectives.
Regarding the code quality, to name a few, 
Bavota and Russo~\citep{bavota2015four} have found that reviewed
code is significantly less likely to contain bugs.
Meanwhile, McIntosh et al.~\citep{mcintosh2016empirical} found that review coverage  shares a significant relationship with software quality.
Kononenko et al.~\citep{kononenko2016code} also reported that the review quality is primarily associated with the thoroughness of the feedback, the reviewer’s familiarity with the code, and the perceived quality of the code itself.
In terms of the review process, Yu et al.~\citep{yu2015wait} explored the
various factors that could impact how long it took for an integrator to merge a pull request. 
Baysal et al.~\citep{baysal2013influence} suggested that these nontechnical factors (e.g., reviewer load and activity, patch writer experience) can significantly impact code review outcomes.
On the other hand, recent researches highlight that code reviews suffer from many challenges, including receiving timely feedback~\cite{macleod2017code}, information needs~\cite{pascarella2018information, wang2021understanding}, and confusion~\cite{ebert2019confusion, ebert2021exploratory}.

\textbf{Continuous Integration.} 
Continuous Integration is a software development practice where team members integrate their work frequently, and each  integration is verified by an automated build (including
test) to detect integration errors~\cite{fowler2006continuous}.
Such automation has been reported to increase productivity significantly in both industrial projects~\cite{goodman2008s, meyer2014continuous, duvall2007continuous}.
Driven by these success stories, the impact of CI on the
open source software development process has become a topic of active research.
On the one side, Hilton et al.~\citep{hilton2016usage} studied the usage of CI in open-source projects and showed evidence that CI helps projects release more often.
Likewise, Vasilescu et al.~\citep{vasilescu2015quality} reported that CI improves the productivity of project teams.
Furthermore, Zhao et al.~\citep{zhao2017impact} extended the impact of CI on other development practices (i.e., the adaptation
and evolution of code writing and submission) and suggested a more nuanced picture of how GitHub teams are adapting to.
On the other side, CI build results are not always a reliable
indication of a code change’s quality, for instance, suffering from brown builds (i.e., build failure that changes to success on at least one build rerun without changing the build setup or source code.).
Several issues have been discovered to cause such build jobs that fail inconsistently, such as  asynchronous calls, multithreading, or test
order dependencies~\citep{gallaba2018noise, ghaleb2019studying}.
Ghaleb et al.~\citep{ghaleb2019empirical} used mixed-effects logistic models to model long build duration across projects and observed that rerunning failed commands multiple times is most likely to be associated with long build duration.
Doriane et al.~\citep{9793972} pointed out that such brown builds not only require multidisciplinary teams to spend more effort interpreting or even re-running the build, leading to substantial redundant build activity, but also slows down the integration pipeline.

\textbf{CI in Modern Code Review.}
Rahman and Roy~\citep{rahman2017impact} reported an exploratory study using
578K automated build entries where they analyzed the impact of
automated builds on the code reviews.
Their results suggested that successfully passed builds are more likely to encourage new code review participation.
To understand how developers use the outcome of CI builds
during code review, Zampetti et al.~\citep{zampetti2019study} empirically investigated the interplay between pull request discussion and the use of CI by means of 64,865 pull request discussions.
They observed that pull requests with passed builds have a higher chance of being merged than failed ones, and pointed out difficulties in properly configuring and maintaining a CI pipeline.
Similarly, Bernardo et al. ~\citep{bernardo2018studying} studied the impact of adopting continuous integration on the delivery time of pull requests and they found that projects deliver merged PRs more quickly after adopting CI.
At the same time, Cassee et al.~\citep{cassee2020silent} observed that with the introduction of CI, pull requests are being discussed less, giving rise to the idea of CI as a silent helper.
Recent work also analyzed the side effects brought by CI.
For instance, Khatoonabad et al.~\citep{khatoonabadi2021wasted} found that being difficult to resolve the CI build  failures is one of the frequent reasons for contributors to abandon their pull requests.
Zhang et al.~\citep{zhang2022pull} reported that, pull requests that need longer CI builds are more likely to take more time for review. 
Moreover, pull
requests which passed the CI builds are more likely to be handled in a shorter time.
Durieux et al.~\citep{durieux2020empirical} empirically analyze restarted and flaky builds on Travis CI and its effect on GitHub pull requests.
Our study further quantifies the patches that are rechecked once and those that are rechecked several times. This allows us to observe justifiable rechecks and potentially wasteful rechecks. These results strengthen the observations of Durieux et al.~\citep{durieux2020empirical}. 
Our work also raises awareness of wasteful rechecks for several stakeholders (e.g., developers, developer experience engineers), such as how the developer workflow and computational resources are impacted by wasteful rechecks.


\section{Conclusion}
\label{sec:conclusion}
Automation has led to efficient processes in software development, especially in the context of the code review and CI processes.
A hidden and opaque by-product generated by these CI test resources can be characterized as software waste.
In this study, we conducted an empirical study on the OpenStack community to explore repeated builds by analysing 66,932 code reviews.
Our results reveal that this waste is prevalent (i.e., 55\%), less likely to change outcomes  (i.e., 42\%), and consumes resources in both waiting and computational times (187.4 years and take 2,200\% more time to review).
Based on this work which has established the extent of the resource waste generated by repeated CI builds, there are many open avenues for future work: understanding how to reduce these wastes by better handling failed CI builds, further studies for better identification  and awareness of such resource waste to developers, and tool support to reduce the waste produced by repeated rebuilds. 


\section{Acknowledgement}
\label{sec:acknowledgement}
This work is supported by Japanese Society for the Promotion of Science (JSPS) KAKENHI Grant Numbers 20K19774, 20H05706, 23K16864, and 21H04877 and JSPS Bilateral Joint Research Project (JPJSBP120239929).
Patanamon Thongtanunam was supported by the Australian Research Council’s Discovery Early Career Researcher Award (DECRA) funding scheme (DE210101091). Yasutaka Kamei was supported by Inamori Research Institute for Science, Kyoto, Japan (InaRIS Fellowship).
We thank the 24 members of the OpenStack community who shared their expert opinions with us.
\balance
\bibliographystyle{ieeetr}
\bibliography{filteredref.bib}

\end{document}

%% file: comments.tex
\usepackage{datenumber}

\definecolor{chestnut}{rgb}{0.8, 0.36, 0.36}

\definecolor{chestnut}{rgb}{0.8, 0.36, 0.36}

\newcounter{dateone}\newcounter{datetwo}%
\newcommand{\daydifftoday}[3]{%
\setmydatenumber{dateone}{\the\year}{\the\month}{\the\day}%
\setmydatenumber{datetwo}{#1}{#2}{#3}%
\addtocounter{datetwo}{-\thedateone}%
\thedatetwo
}